\documentclass[prd,amsmath,amssymb,nofootinbib,floatfix,superscriptaddress]{revtex4}

\usepackage{graphicx,bm,xcolor, bbold}
\usepackage{enumerate}
\usepackage{graphicx,bm}
\usepackage[normalem]{ulem}

\begin{document}


%
%

\title{Gauge-invariant field strengths in $QCD$}

\author{Adriano Di Giacomo}

\affiliation{Dipartimento di Fisica Universita' di Pisa and I.N.F.N. Sezione di Pisa\\
3 Largo B. Pontecorvo 56127 Pisa Italy
\\
adriano.digiacomo@df.unipi.it}

\begin{abstract}
Gauge-invariant field strengths, defined as parallel transports to infinity of ordinary field strengths, naturally emerge in  a few  physical phenomena governed    by                                                                              $QCD$. One of them is confinement of colour. Despite the arbitrariness in their definition coming from the freedom in the choice of the path of the parallel transport to infinity, the request of differentiability with respect to the position $x$ strongly constrains  their correlation functions. Strong constraints also come from translation and Lorentz invariance. Gauge invariant field strengths also appear in the non abelian Stokes theorem, and allow to understand basic properties of the vacuum by use of lattice data.

  \vspace{.3cm}
  
 Keywords : Confinement ; Field theory ; QCD ; Lattice QCD.
\end{abstract}

\maketitle


\section{Introduction}	
The idea of constructing gauge invariant charged fields  by parallel transport to infinity    goes back
to Dirac   \cite{Dirac}. Dirac's  program was to formulate $QED$ in terms of gauge-invariant quantities. 

Gauge-invariant field strengths $F_{\mu \nu}(x)$   in $QCD$ naturally emerge in the construction of the creation operator of monopoles  $\mu$ \cite{digiacomo1,digiacomo2}. $\langle \mu \rangle$ is the order parameter for the condensation of monopoles in the vacuum \cite{dig}, which  can be  the mechanism responsible for  confinement of colour \cite{'tH,m}.

 $F_{\mu \nu}(x)$ is defined as \cite{digiacomo1}
\begin{equation}
F_{\mu \nu} (x) =  V_C (x, \infty) G_{\mu \nu}(x)V^{\dagger}_C (x,\infty) \label{gif}
\end{equation}
Here
\begin{equation}
G_{\mu \nu}(x) =  \partial_{\mu} A_{\nu} (x) - \partial_{\nu} A_{\mu}(x) + ig [A_{\mu}(x),A_{\nu}(x)]
\end{equation}
is the field strength tensor in the standard notation. For the sake of definiteness  we shall assume that the gauge group is $SU(N)$, and we shall work in the fundamental representation where both $A_{\mu}(x)$ and $G_{\mu \nu}(x)$ are $N \times N$ traceless matrices.
$V_C (x,\infty)$ is the parallel transport from $x$ to $\infty$ along a continuous  line $C$ which is  a priori arbitrary. 
\begin{equation}
V_C (x,\infty)= P \int _{x \hspace{.1cm}C}^{\infty }  \exp (i A_{\mu}(x)) dx_{\mu}
\end{equation}
$F_{\mu \nu}$ depends on $x$ but also on the infinitely many variables which identify the path $C$ and the field  $A_{\mu}$ on it.

The operator $\mu ( \vec x, t)$ which creates a monopole at the point $\vec x$ and time $t$ is \cite{dig}
\begin{equation} 
 \mu(\vec x , t) = exp  \big ( \int d^3 y  G_{0 i}^3 (\vec y, t ) \frac{1}{g}  A_{\perp}^i (\vec y -\vec x) \big ) \label{mu}
\end{equation} 
Here $\frac{1}{g}\vec A_{\perp} (\vec z) $ is the classical field of a monopole \cite{'tH1}  \cite{Pol} in the transverse gauge $\vec \nabla \vec A_{\perp} =0$. $ G_{0 i}^3 $ is the $i-th$ spatial component of the electric field strength operator in the colour direction where the monopole lives ( Abelian Projection), which we conventionally denote by the index 3. In the convolution of Eq(\ref{mu}) only its transverse part survives which is  the canonically conjugate momentum to the field $ A_{\perp}^{3i} $.  As a consequence, if we denote by
$| A^{3 i}_{\perp} (\vec z) \rangle$ a state of the transverse field in the Schrodinger  representation at time $t$,
\begin{equation}
\mu ( \vec x, t) | A^{3 i}_{\perp} (\vec z) \rangle = | A^{3 i}_{\perp} (\vec z) +\frac{1}{g}  A_{\perp}^i (\vec z -\vec x)\rangle
\end{equation}

showing that the operator $\mu (\vec x,t)$ does create a monopole.

If the colour direction 3 is that of  a generator of a local $SU(N)$ group as suggested in Ref\cite{'tH2} and as usually happens in lattice simulations\cite{dlmp} $\mu$ breaks the local symmetry $SU(N)$ to $SU(N)/U(1)$.  As a consequence the vacuum expectation value $ \langle \mu \rangle$ does not exist because of Elitzur's theorem \cite{El}, which states the impossibility of spontaneously breaking local symmetries. This is the origin of the troubles in computing $\langle \mu \rangle$ \cite{dlmp} \cite{cddlp} \cite{bcdd}. Of course these problems do not exist in  $U(1)$ gauge theory where the electric field is gauge invariant \cite{dp}.  

The way out is simple and physical \cite{digiacomo1} \cite{digiacomo2}: make the operator $\mu$ gauge-invariant by parallel transport of  $T^3$ to infinity. The subgroup in which monopole lives becomes in this way gauge invariant. 

This amounts to replace $G_{\mu \nu}$ by $F_{\mu \nu}$ in Eq(\ref{mu}). 
\begin{equation}
 \mu(\vec x , t) = exp  \big ( \int d^3 y  F_{0 i}^3 (\vec y, t ) \frac{1}{g}  A_{\perp}^i (\vec y -\vec x) \big ) \label{mu'}
\end{equation}

 The choice of  the path $C$ is  arbitrary.

  However, if we want to differentiate $F_{\mu \nu}(x)$ with respect to $x$,  we have to keep the parameters which describe $C$ fixed:
  
   $F_{\mu \nu}(x+\delta x_{\rho}) =  V_C (x+\delta x_{\rho}, \infty) G_{\mu \nu}(x+ \delta x_{\rho})V^{\dagger}_C (x+ \delta x_{\rho},\infty)\approx F_{\mu \nu}(x) + \hspace{.2cm}\delta x_{\rho} \partial_\rho F_{\mu \nu}(x)$. 
   
 If $ l_{\rho} (x) \approx (1 + iA_{\rho}(x) \delta x_{\rho})$ is the link joining $x$ to $x+\delta x_{\rho}$,  

$V_C (x+ \delta x_{\rho},\infty) \approx  V_C (x,\infty)l_{\rho}^{\dagger}\approx V_C (x,\infty)(1 - iA_{\rho}(x) \delta x_{\rho})$,

  $G_{\mu \nu}(x+ \delta x_{\rho}) \approx G_{\mu \nu}(x) + \partial_\rho G_{\mu \nu} (x) \delta x_{\rho}$ and by use of simple algebra
\begin{equation}
\partial_{\rho} F_{\mu \nu}(x) = V_C(x,\infty) D_{\rho} G_{\mu \nu}(x) V^{\dagger} _C (x, \infty) \label{der}
\end{equation}
$D_{\mu}  = \partial_{\mu} - ig[A_{\mu}, \hspace{.15cm}  ]$ is the covariant derivative.
In particular, for $\mu =\rho$
\begin{equation}
\partial_{\mu} F_{\mu \nu} = V_C (x,\infty) D_{\mu} G_{\mu \nu}(x) V ^{\dagger}_C (x, \infty) \label{CEM}
\end{equation}
In absence of quarks $D_{\mu} G_{\mu \nu}=0 $ and

\begin{equation}
\partial_{\mu} F_{\mu \nu} = 0 \label{div}
\end{equation}
Eq.(\ref{div}) has been used in Ref.'s \cite{digiacomo1}, \cite{digiacomo2}  to prove that infrared divergences in the definition of the order parameter exactly cancel to all orders of strong coupling expansion, so that $\langle \mu \rangle$ is well defined. 

Differentiability of $F_{\mu \nu}$ strongly constrains the choice of the path $C$: two neighbouring sites must have the same $C$ except for the small displacement separating them. This implies by continuity that the paths $C$ $C'$ of any pair of fields  $F_{\mu \nu}(x)$ $ F_{\rho \sigma}(x')$ overlap from some point on in their way to infinity. Whatever choice of the paths $C$ requires by differentiability that they all meet in a point and from it jointly proceed to infinity.

  The replacement  of $G_{\mu \nu}(x)$ by $F_{\mu \nu}(x) $ provides  the solution of an old standing problem  \cite{Greensite}  \cite{Dmo}. In any local abelian projection flux tubes joining a confined $q-\bar q$ pair should be oriented in colour space in the direction of the effective Higgs field of the monopole. Lattice simulations instead show that their
  orientation is compatible with random \cite{Greensite} \cite{Dmo}. This fact was indicated as a difficulty of the mechanism of dual superconductivity in explaining colour confinement \cite{Greensite}. With  the gauge invariant order parameter Eq.(\ref{mu'})  instead the colour orientation of the monopole is fixed at infinite distance and  memory of it gets lost in the parallel transport from  infinity to the points inside the flux tube.
  
  Gauge invariant field strengths naturally   appear also in the non abelian Stokes theorem  \cite{Aref'}
  \begin{equation}
  P \exp \big (\int_{\partial \Sigma} A_{\mu} dx^{\mu} \big ) =  P \exp \big (\int _{\Sigma} F_{\mu \nu}(x_1,x_2) {dx_1}^{\mu} {dx_2}^{\nu} \big ) \label{nast}
  \end{equation}
$\Sigma$ is a surface e.g. a Wilson loop , $\partial\Sigma$ its border. $P$ in the right side of Eq(\ref{nast})  means ordered product in one of the two coordinates, say  ${x_1}^{\mu}$.The path $C$  in the definition of $F_{\mu \nu}$ in Eq (\ref{nast}) has a special form: the point $O$ is chosen at the origin and the parallel transport as the product of subsequent parallel transports along the coordinate axes from $0$ to $x_i$ in any order  \cite{Aref'}.
Non abelian Stokes theorem is relevant e.g. to the model of stochastic vacuum of Ref.'s (\cite{dosch} \cite{simonov} \cite{ddss}).

In Section 2 we  give the general definition of vacuum correlations   of gauge invariant field strengths  and discuss its dependence on the path $C$ which appears in their definition Eq(\ref{gif}).We do that in the strong coupling expansion of the theory as presented in \cite{creutz1}, \cite{creutz2}. We then compute the one-point and the two point function as examples..

In Section 3 we compute the string tension from lattice data by use of the non abelian Stokes theorem. The area term of the Wilson loop  measured by lattice simulations is compared to its value computed by use
of the non abelian Stokes theorem plus the assumption at the basis of the stochastic vacuum model of  Ref.'s (\cite{dosch} \cite{simonov}) that gauge invariant two point functions provide a good approximation to physics in a cumulant expansion.  These correlations are directly extracted  from lattice simulations \cite{dP}.  The result is a direct quantitative test of the 
validity of the assumption.

\section{Vacuum correlations of $F_{\mu \nu}(x)$.}
 On the lattice  the field strength is 
\begin{equation}
G_{\mu \nu}(n) \approx i \Im \Pi_{\mu  \nu} (n)
\end{equation}
$\Im$ denotes the Imaginary (anti-hermitian) part.
 $\Pi_{ \mu \nu}$ is the open plaquette at $n$ in the hyperplane $ \mu \nu$. Its non singlet part   is the lattice  field strength whose  parallel transport to infinity is the lattice gauge-invariant field strength.
 \begin{equation}
 F_{\mu \nu}(n) = V_C (n, \infty)  \Im \Pi_{\mu \nu}(n)V^{\dagger}_C (n,\infty)
 \end{equation}
 
 The m-point function is defined as
 \begin{eqnarray}
\langle F _{\mu_1 \nu_1} (n_1 )  F _{\mu_2  \nu_2} (n_2 ).... F _{\mu_m  \nu_m} (n_m )\rangle  =  \frac{1}{\int \Pi dl \exp(-\beta S)} . \nonumber \\ \int \Pi dl  \exp(-\beta S)\frac{1}{N} Tr [ F _{\mu_1 \nu _1} (n_1) F _{\mu_2  \nu_2} (n_2 ).... F _{\mu_m \nu_m} (n_m )] \hspace{.25cm} \label{act}
\end{eqnarray}

The notation is standard. \hspace{.2cm}$\beta S$ is the action , \hspace{.1cm}
$S= \sum_{n , \mu \nu} \Re (1 -  \Pi_{\mu \nu}(n))$, \hspace{.3cm} $\Re$ denotes real (hermitian ) part,

  \hspace{.1cm} $\beta=\frac{2N}{g^2}$ \hspace{.1cm} and\hspace{.1cm} $g$  is the $QCD$ coupling constant.

The system is defined on a lattice 
and the sum is extended to all sites $n$ and orientations $\mu \nu$. The integration is performed on all the elementary links $l$ parallel to the coordinate axes. 
These links are elements of the gauge group in the fundamental representation. Since all physical quantities are written on the lattice as products of such links
all observables are integrals on the group of products of elementary links. Such integrals are well defined if the group is compact (See e.g. Ref. \cite{creutz2}).

The strong coupling expansion of Eq(\ref{act}) is obtained by expanding the exponentials in powers of $\beta\propto \frac{1}{g^2}$.
The integral on each link is extended to the group and exists if the group is compact. The normalisation can be fixed as $\int dl =1$.
As a consequence the links which do not appear in a given  term of the expansion of $\exp(-\beta S) \Pi _k (F^{i_k} _{\mu _k \nu _k} (x_k) )$ or of $\exp(-\beta S)$ contribute a factor of $1$ and can be ignored. 

To understand the structure of the correlation functions Eq(\ref{act}) we re-wright the gauge invariant fields in terms of the ordinary field strengths and of the parallel transports $V_C(x,\infty)$, keeping into account that differentiability of the fields $F_{ \mu \nu}(n)$ implies 

\begin{equation} 
V_{C_i} (n_i, \infty) = V (O,\infty) V_{c _i}(n_i, O) .  \label{fract}
\end{equation}
$c _i $ is the part of the path $C_i$ going from $n_i$ to the point $O$ from which on all the parallel transports coincide and are equal  to $V(O,\infty)$.
Each field $F_{\mu_i \nu_i} (n_i) $ can then be written in the form 
\begin{equation}
F_{ \mu _i \nu _i}(n _i) = V(O, \infty)  V_{c _i }(n _i, O)G_{\mu _i \nu _i}(n _i) V^{\dagger}_{c_i } (n _i, O)V^{\dagger}(O, \infty) 
\end{equation}

$G_{\mu \nu}(n)$ is a linear combination of the generators  $T^i$ of the gauge group. The same is true of $F_{\mu \nu}(n)$ which is a unitary transform   of $G_{\mu \nu} (n)$. We can then write

 \begin{equation}
 F_{\mu \nu}(n) = K \sum_i T^i F^i _{\mu \nu}(n)
\end{equation}

where $K$  is a constant which fixes the normalisation of $F^i _{\mu \nu}(n)$. 

From the relation 
\begin{equation}
Tr[T^a F_{\mu \nu}(n) ] = \delta_{ai} Tr[(T^i)^2] K F^a _{\mu \nu}(n)
\end{equation}

By choosing $ K=\frac{1}{Tr[(T^i)^2]}$   we have
\begin{equation}
F^i _{\mu \nu} (n) =Tr[ T^i F_{\mu \nu}(n)]  
\end{equation}

The correlation function Eq(\ref{act})  reads then

\begin{equation}
\langle F_{\mu_1 \nu_1}(n_1)...F_{\mu_m \nu_m}(n_m)\rangle  =K^m \frac{1}{N} Tr [ T^{i_!}....T^{i_m}] \langle F^{i_1} _{\mu_1 \nu_1} (n_1).......F^{i_m} _{\mu _m \nu_m}(n_m) \rangle \label{cfu}
\end{equation}
Explicitly  each factor $F^i _{\mu \nu} (n)$ in Eq(\ref{cfu}) can be written in terms of the fields $G_{\mu \nu}(n)$ in the form

$F^i_{\mu \nu} =Tr[ T^i V_C (\infty, n) G_{\mu \nu}(n)V^{\dagger}_C(\infty ,n)]$, or by use of Eq (\ref{fract})
\begin{equation}
F^i_{\mu \nu} =Tr[ T^i V (\infty, O) V_c (O ,n)G_{\mu \nu}(n)V^{\dagger}_c (O,n)V^{\dagger}(\infty ,O)]
\end{equation} 
$V_c (O ,n)$  depends on the point $n$ and is the part of the path $C$ going from $n$ to the point $O$ where all the paths merge. $V(\infty, O) $ is independent of the point $n$.

\vspace{.2cm}

Let us consider first the correlation function Eq(\ref{act}) at the leading  (zeroth) order in the strong coupling expansion \hspace{.2cm}( $\beta=0$ ). The only non trivial integrals are those on the links of $V_C (O, \infty)$ and those on  the links of the fields  $ G_{\mu _i \nu _i }( n_i )$
 and of the parallel transports $V_{c_i } (n _i, O)$. Let us first perform the integrals on the links of $V(O, \infty)$
 
There is a group theoretical property which simplifies the calculation of such integrals.

As a consequence of compactness for any arbitrary function $f(l)$ and for any two    group elements  $g_1$ $g_2$\cite{creutz2}
 \begin{equation}
 \int dl f(l) = \int dl f (g_1 l g_2)   \label{invprop}
\end{equation}
 
Our integrand is a function  of $V(O, \infty)$ which is the product of all the links of the parallel transport from $O$ to $\infty$ say $l_1 l_2 ...l_{k-1} l_k l_{k+1}...$. Calling $g_1 = l_1 l_2... l_{k-1}$  and $g_2 = l_{k+1}l_{k+2}...$which  are both  elements of the compact group, the integration on the links of  $V_C (O, \infty)$ reduces, by virtue of Eq(\ref{invprop}), to the integration on the one variable $l_k$ the others giving each as a factor the volume of the group which is $=1$. Integrating on all of them gives the same result as integrating on one of them, independent on its position and orientation in space.  For the sake of simplicity  we shall  denote by  $W$ the link left to be integrated over.  

As  a result  we get 
\begin{eqnarray}
\langle F _{\mu_1 \nu_1} (n_1 )  F _{\mu_2  \nu_2} (n_2 ).... F _{\mu_m  \nu_m} (n_m ) \rangle  =  \int \Pi dl  \int dW  K^m \hspace{1.cm}    \nonumber \\ \frac{1}{N} Tr[T^{i_1} .. T^{i_m}]
  \hspace{.1cm}  \Pi _{l=1..m}  Tr  [ W^{\dagger} T^{i_l} W V_{c_l} (O,n_l)G_{\mu _l \nu_l}(n_l) V^{\dagger} _{c_l}(O, n_l)]
\end{eqnarray}
The integral on $W$ can be performed and only depends on the gauge group \cite{creutz1} \cite{creutz2}. We define $I_{i_1..i_m, j_1..j_m, k_1...k_m, l_1..l_m }$ as
\begin{equation} 
I_{i_1..i_m, j_1..j_m, k_1...k_m, l_1..l_m } \equiv \int dW W_{i_1 j_1} W^{\dagger} _{k_1 l_1} .....W_{i_m j_m} W^{\dagger} _{k_m l_m}
\end{equation}

  The integral is extended to the group and $W$ denotes  the generic group element in the fundamental representation.
A few  such integrals are known from the strong coupling expansion of the theory Ref (\cite{creutz1} , \cite{creutz2}).  In general $ I_{i_1..i_m, j_1..j_m, k_1...k_m, l_1..l_m } $ is a sum of products of delta's of indexes $i, k$ times products of deltas in the indexes   $j, l$. \cite{creutz1}  For values of $m \ge 3$ also full antisymmetric invariant tensors in $i k$ and $j l$ indexes are present\cite{creutz1} \cite{creutz2}.

After integration on $W$ the correlation becomes
\begin{eqnarray}
\langle F _{\mu_1 \nu_1} (n_1 )  F _{\mu_2  \nu_2} (n_2 ).... F _{\mu_m  \nu_m} (n_m ) \rangle  =  \int \Pi dl   K^m    \frac{1}{N} Tr[T^{i_1} .. T^{i_m}] \nonumber \\
  I_{i_1..i_m, j_1..j_m, k_1...k_m, l_1..l_m }   T^1 _{i_1 k_1} (G^O  _{\mu_1 \nu_1} (n_1))_{j_1 l_1}. ... T^m _{i_m k_m} (G^O _{\mu _m \nu _m}) _{j_m  l_m}(n_m) \label{corr}
\end{eqnarray}
 By $G^O_{\mu \nu}(n)$ we have denoted the parallel transport along $c$ of  $G_{\mu \nu} (n)$ from the point $n$ to the point $O$ where all the paths $C$ meet .
 \begin{equation}
 G^O _{\mu \nu}(n) = V_c (O,n) G_{\mu \nu} (n) V^{\dagger} _c (O,n) 
 \end{equation}
 For $m=1$
\begin{equation}
I_{i_1,j_1, k_1,l_1}  = \frac{1}{N} \delta_{i_1 k_1} \delta_{j_1 l_1}
\end{equation}
the point $O$ can be taken equal to $n$
and
\begin{equation} 
\langle F_{\mu \nu} (n) \rangle = K (\frac{1}{N} Tr[T^i] ) Tr [T^i]  Tr[G_{\mu \nu} (n)] =0 \label{f1}
\end{equation}
$\langle F_{\mu \nu} (n) \rangle$  is zero to all orders in the strong coupling expansion because of  the trace factor in front. Also the other two  factors are zero separately but they could be affected by higher orders in the strong coupling expansion. This is an expected result due to the fact that vacuum is a colour singlet.

For $m=2$ \cite{creutz1} \cite{creutz2}

\begin{eqnarray}
I_{i_1 i_2,j_1,j_2,l_1 l_2, k_1 k_2}= \frac{1}{N^2 -1}[ (\delta_{i_1 k_1} \delta_{i_2 k_2} \delta_{j_1 l_1} \delta_{j_2 l_2} + \delta_{i_1 k_2} \delta_{i_2 k_1} \delta_{j_1 l_2} \delta_{j_2 l_1}) \nonumber \\ -\frac{1}{N} (\delta_{i_1 k_1 }\delta_{i_2 k_2} \delta_{j_1 l_2}. \delta_{j_2 l_1} + (\delta_{i_1 k_2} \delta_{i_2 k_1} \delta_{j_1 l_1} \delta_{j_2 l_2})] \label{I2}
\end{eqnarray}
Inserting this expression in Eq(\ref{corr}) with $m=2$ gives
\begin{eqnarray}
\langle F_{\mu_1 \nu_1}(n_1) F_{\mu_2 \nu_2 }(n_2) \rangle &= &\int \Pi dl  (N^2-1) K^2  Tr [(T^i)^2]^2 (N^2 -1)^{-1} \frac{1}{N} Tr[ G^O _{\mu_1 \nu_1}(n_1) G^O _{\mu_2 \nu_2}(n_2)] \nonumber \\
&=&  \int \Pi dl \frac{1}{N} Tr [ G_{\mu_2 \nu_2}(n_2) U_{21} G_{\mu_1 \nu_1}(n_1)U_{21}^{\dagger} ] \label{int}
\end{eqnarray}
Here $U_{21} = V^{\dagger} _{c_2} (n_2 , O) V _{c_1} (n_1, O) $ is the parallel transport from $n_1$ to $n_2$   along the path $c_1 - c_2$
Only the second term  of the integral Eq(\ref{I2}) contributes. The first term and the third term give a contribution proportional to $Tr[ T^1] .Tr[T^2] $ which is zero. The fourth term gives 
a contribution proportional to $Tr [G_{\mu_1 \nu_1} (n_1)] Tr[G_{\mu_2 \nu_2} (n_2) ]$which is zero again. The factor $N^2-1$ in the numerator comes from the sum on the generators which are $N^2 -1$ for the gauge group $SU_N$, the same factor in the denominator comes from the integral Eq(\ref{I2}). The factor $K^2 $ cancels with $(Tr [(T^i)^2])^2$

This two point function coincides formally with the gauge invariant two point function of the stochastic vacuum model \cite{dosch} \cite{simonov}. $O$ is the reference point in their language, $c_1, c_2$ the paths connecting $n_1, n_2$ to it.
In the stochastic vacuum model  the two point function Eq(\ref{int}) is the basic building block of the theory and is  defined at any value of $\beta$. Of course the choice of the point $O$ as well as the paths  $C_1, C_2$ are arbitrary.

Correlations like the one in Eq(\ref{int})  have been studied on the lattice in Ref [ \cite{dP}  \cite{ddss}] for paths $c_1 - c_2$ a straight line. We shall come back to them in the next section.

In the gauge invariant field approach at non zero $\beta$ Eq(\ref{act}) we are not allowed in general to perform the integration on the line $C(O, \infty)$ before integrating on the other links. Configurations exist in which links of the plaquettes of the action are superimposed to it.  However one can argue that their contribution is negligible so that the correlation function is the  average of the operator Eq(\ref{int}) weighted with the action, or
\begin{equation}
\langle F_{\mu_1 \nu_1}(n_1) F_{\mu_2 \nu_2}(n_2) \rangle = \frac{\int \Pi dl \exp(- \beta S) \frac{1}{N} Tr[G_{\mu_1 \nu_1}(n_1) U_{21} G_{\mu_2 \nu_2} (n_2)U^{\dagger} _{21}]} {\int \Pi dl \exp (- \beta S)} \label{tpf}
\end{equation}

 The reason is that in the average Eq(\ref{act}) only connected configurations contribute to the possible deviation from Eq(\ref{tpf}). The contributions of disconnected configurations cancel between numerator and denominator. At first order in $\beta$ the possible contributions come from one plaquette terms in the expansion of $\exp(- \beta S)$ , but they vanish because the phase integral of a single link vanishes. For a non zero contribution one needs a term $O(\beta^2) $ with two plaquettes superimposed. To have it connected at least one of the edges must lie on the curve $C(O, \infty)$. The integral on the other edges forces by gauge invariance its contribution to be the same   they would give in absence of $C(O, \infty)$, so that the configuration is in fact disconnected and its contribution cancels. Non zero contribution to deviations from  the equation come from higher orders in the expansion and can be neglected.
 
 The only memory left of the parallel transports to $\infty$ is in the path of the parallel transport $U_{12}$. A special case is when the line $c_1 - c_2$ is  a straight line \cite{dP}.

We will not discuss here correlation functions with $m \ge 3$. We only notice that Eq(\ref{corr}) can be of help in understanding their colour structure.

\section{Computing the string tension}
In this section we shall use the lattice results of Ref(\cite{dP}) to compute the string tension by use of the non abelian Stokes theorem \cite{Aref'}. As a contour we shall use a rectangular Wilson loop $w$ in the $(x,t)$ plane, of size $R,T$. 
\begin{equation}
w = P \int _{\partial w}\exp (i \int A_{\mu} dx^{\mu})
\end{equation}
It is well known that,  as a consequence of confinement, at large $R, T$ $w$ obeys the area law
\begin{equation}
\langle w \rangle\propto \exp(- \sigma RT)
\end{equation}
$\sigma $ is the string tension, the attractive force between a pair $q \bar q$ of static quarks at large distances. $\sigma$ is measured on lattice. For $SU(3)$ gauge theory \cite{creutz2} the lattice spacing is 
\begin{equation}
a=\frac{1}{\Lambda} (\frac{8}{33} \pi^2 \beta)^{\frac{51}{121}}\exp(-\frac{4}{33}\pi^2 \beta) [1+O(\frac{1}{\beta})] \label{RG}
\end{equation}
 as dictated by renormalisation group at two loops in perturbation theory.
  Lattice simulations relate the scale $\Lambda$ to the string tension \cite{creutz2}
\begin{equation}
\Lambda =(6.0 \pm 1.0) 10^{-3} \sigma ^{\frac{1}{2}}
\end{equation}
For sure more accurate determinations  exist than this pioneering result, but it is not easy to find them in the literature. 

The non abelian Stokes theorem states that
\begin{eqnarray}
\langle w \rangle = \langle P \exp( i \int^T _0  dx^0 \int ^R _0 dx^1 F_{ 1 0}(x^0 ,x^1)\rangle \approx \nonumber\\
1 - \int^R _0 dx^1 _1 \int ^{x^1_1} _0 dx^2 _1 \int ^T _0 dx^1_0 \int^{T} _0 dx^2 _0 \langle F_{01}(x^1_1, x^1 _ 0) F_{01}(x^2 _1, x^2_0) \rangle + ... \label{St}
\end{eqnarray}
The linear term in $F_{\mu \nu} $ vanishes since $\langle F_{\mu \nu}(n) \rangle =0$ , Eq(\ref{f1}). 

Higher correlations are indicated by dots in Eq(\ref{St}).  In the spirit of the Stochastic Vacuum Model \cite{dosch} \cite{simonov} the two-point functions dominate and the second  order term can be considered as the first term in a cumulant expansion. This means
\begin{equation}
 \sigma = \frac{1}{RT}\int ^R _0 dx^1_1 \int^{x^1_1}_0 dx^2_1 \int^T_0 dx^1_0 \int^T _0 dx^2 _0 \langle \frac{1}{N} Tr [  G_{01}(x^2 _1, x^2_0)U_{12}G_{01}(x^1_1, x^1 _ 0)U^{\dagger} _{12} ]\rangle \label{nast1}
\end{equation}
By invariance under translations the quantity $ \Phi (z) =\langle \frac{1}{N} Tr [  G_{01}(x^2 _1, x^2_0)U_{12}G_{01}(x^1_1, x^1 _ 0)U^{\dagger} _{12} ]\rangle$ only depends on the difference $ z \equiv x^2 -x^1 $ and not on $x^1$. The integral on $x^1$ can be computed and just cancels $RT$ in the denominator of Eq(\ref{nast1}).

The field  $F_{\mu \nu}$ in Ref(\cite{Aref'}) is defined in such a way that   the parallel transport $U_{12}$ between the two  points $n_1, n_2$ is along a straight line. Strictly speaking this is true when the two points have the same value  for one of the coordinates, say $z^1$ . However by symmetry properties of Eq(\ref{tpf}) this is true in general: indeed the quantity $\Phi (z)$ transforms as a tensor under Lorentz transformations and the transform of a straight line  parallel transport is again straight.    Gauge invariant correlations become covariant under Lorentz transformations and translations when the parallel transport $U_{12}$ is a straight line. There exist two  invariant functions  $D(z)$ and $D_1 (z) $ of  $z_{\mu}$ \cite{dosch} \cite{simonov}. \cite{dP}
 
 In particular for the correlation of two components of the chromo-electric field we have
 
 \begin{equation}
\Phi(z)= \frac{1}{N} Tr [  G_{0i}(x_2 )U_{12}G_{0j}(x_1)U^{\dagger} _{12} ] = \delta_{ij} \big ( D(z) +D_1 (z) + [z_0]^2  \frac{d D_! (z)}{d z^2} \big ) + z_i z_j \frac{d D_! (z)}{d z^2}
 \end{equation}
  \vspace{.2cm}
 For the two point function in Eq(\ref{nast1}) since $z$ has only the components $0$ ,$1$ we have 
 
\begin{equation}
  \frac{1}{N} Tr [  G_{01}(x^2 _1, x^2_0)U_{12}G_{01}(x^1_1, x^1 _ 0)U^{\dagger} _{12} ] =   D(z) + D_1 (z) + z^2 \frac{d D_! (z)}{d z^2} 
  \end{equation}
  The functions $D(z)$ and $D_1 (z)$ have been determined\cite{dP} by lattice simulations of quenched $SU(3)$   gauge theory. They are decreasing exponentials of $\rho \equiv \sqrt(z^2 )$ with the same slope $\frac{1}{\lambda}$
  \begin{equation}
  D(z) = A \exp(- \frac{\rho}{\lambda}) \hspace{1.cm}   D_1 (z) = A_1 \exp(- \frac{\rho}{\lambda})
  \end{equation}
  with \hspace{.5cm} $\frac{A}{\Lambda^4}=3.6 \times 10^8 $  , \hspace{2.0cm}  $\frac{A_1}{\Lambda^4}=1.25 \times 10^8 $, \hspace{2.0cm} $\frac{1}{\lambda  }  = 183 \Lambda $.  
   \vspace{.5cm}
  
 The error is 
  smaller  than the unity in the last  figure. The scale $\Lambda$ is the same as  that of Eq.(\ref{RG}).
  
  Let us now compute $\sigma$
  If the length $\lambda$ is $<< R,T $, the integral can be extended to $\infty$ and
  \begin{equation} 
  \sigma = \pi \int^{\infty} _0 \rho d \rho[ D(\rho) + D_1(\rho) +\frac{\rho}{2} \frac{d D_1}{d \rho}] \label{dd1}
  \end{equation}
  
  The two terms in $D_1$ cancel each other and the final result is
  \begin{equation}
  \sigma = \pi A \lambda ^2   \label{fin}
  \end{equation} 
  The accuracy at which Eq(\ref{fin}) is satisfied is the accuracy  at which Eq(\ref{St}) i.e. the stochastic model can provide a good approximation to $QCD$. 
  
  Numerically
  $\frac{\sigma}{\Lambda^2}  = \pi \frac{A}{\Lambda^4}\lambda^2 \Lambda^2$, or
  \begin{equation}
  3,4 =  2,77\pm .90
  \end{equation}
  A better determination of the string tension would give a better estimate of the approximation of the stochastic vacuum model.
  
  A final general remark on gauge invariant field strengths. They naturally appear in some phenomena governed by  $QCD$. They are not local fields in the sense of field theory but can provide a field theoretical language for models based on gauge invariant correlations.

\end{document}